\documentclass[a4paper,10pt]{revtex4}
\usepackage{graphicx} 
\usepackage{amsmath} 
\usepackage{amssymb} 
\usepackage{bm} 
\usepackage{dcolumn}
\usepackage{color}
\usepackage{mathrsfs}
\usepackage{amsfonts}
\usepackage{varioref}
\usepackage{mathrsfs}
\usepackage{graphicx}
\usepackage{latexsym}
\usepackage{amsmath}
\usepackage{amssymb}
\usepackage{textcomp}
\usepackage{amsbsy}
\usepackage{graphics}
\usepackage{epstopdf}
\usepackage{color}
\usepackage[caption=false]{subfig}

\RequirePackage[colorlinks,citecolor=blue,urlcolor=magenta,linkcolor=blue]{hyperref}
\input epsf

\allowdisplaybreaks[4]
\newcommand{\e}{\mathrm{e}}

\begin{document}

\tolerance=5000

\title{{
Modified cosmology from the thermodynamics of apparent horizon}}

\author{Shin'ichi~Nojiri$^{1,2}$\,\thanks{nojiri@gravity.phys.nagoya-u.ac.jp},
Sergei~D.~Odintsov$^{3,4}$\,\thanks{odintsov@ieec.uab.es},
Tanmoy~Paul$^{5}$\,\thanks{pul.tnmy9@gmail.com}} \affiliation{
$^{1)}$ Department of Physics, Nagoya University,
Nagoya 464-8602, Japan \\
$^{2)}$ Kobayashi-Maskawa Institute for the Origin of Particles
and the Universe, Nagoya University, Nagoya 464-8602, Japan \\
$^{3)}$ ICREA, Passeig Luis Companys, 23, 08010 Barcelona, Spain\\
$^{4)}$ Institute of Space Sciences (ICE, CSIC) C. Can Magrans s/n, 08193 Barcelona, Spain\\
$^{5)}$ Department of Physics, Chandernagore College, Hooghly - 712 136, India.}


\tolerance=5000

\begin{abstract}
In the realm of the Bekenstein-Hawking entropy, the thermodynamics of apparent horizon bridges with the usual FLRW 
(Friedmann-Lema\^{i}tre-Robertson-Walker) equation 
only for a special case where the matter field is given by a perfect fluid having equation of state (EoS) parameter $= -1$, i.e 
$p = -\rho$ with $\rho$ and $p$ represent the energy density and the pressure of the fluid, respectively. 
To include the case $p \neq -\rho$, we consider the modification of the Bekenstein-Hawking entropy in the present work. 
In particular, we develop an entropy function that leads to the usual FLRW equations, for a $general$ EoS of the matter fluid given by $p = w\rho$, directly 
from the thermodynamics of the apparent horizon. 
The newly developed entropy acquires a correction over the Bekenstein-Hawking entropy and differs from the known entropies 
like the Tsallis, R\'{e}nyi, Barrow, Sharma-Mittal, Kaniadakis, and Loop Quantum Gravity entropies proposed so far. 
Based on this finding, we examine how the Friedmann equations of the apparent horizon cosmology are accordingly modified 
if one starts with a general entropy depending on the Bekenstein-Hawking entropy. 
This results in some interesting cosmological consequences during the early and late stages of the universe. 
\end{abstract}

\maketitle

\section{Introduction}\label{SecI}

With the discovery of the Hawking radiation, it has been clarified that a black hole is a thermal object with entropy that is proportional to 
the area of the horizon and called the Bekenstein-Hawking entropy \cite{Bekenstein:1973ur, Hawking:1975vcx}, 
which strongly suggests the relation between gravity and thermodynamics \cite{Jacobson:1995ab, Padmanabhan:2003gd, Padmanabhan:2009vy, Hayward:1997jp}. 
Not only to the event horizon of the black hole, by applying the thermodynamics to the Hubble horizon in cosmology, but it has also been shown that 
the FLRW equations can be obtained from the first law of thermodynamics \cite{Cai:2005ra,Akbar:2006kj,Cai:2006rs,Akbar:2006er,Paranjape:2006ca,
Sheykhi:2007zp,Jamil:2009eb,Cai:2009ph,Wang:2009zv,Jamil:2010di,Gim:2014nba,Fan:2014ala,DAgostino:2019wko,Sanchez:2022xfh}.

On the other hand, under several situations, other kinds of entropies that are different from the Bekenstein-Hawking entropy have been discussed. 
Such entropies are the Tsallis entropy \cite{tsallis} (see also \cite{Ren:2020djc,Nojiri:2019skr}), 
the R{\'e}nyi entropy (\cite{renyi}, see also \cite{Nishioka:2014mwa, Czinner:2015eyk,Tannukij:2020njz,Promsiri:2020jga,Samart:2020klx}), 
the Sharma-Mittal entropy \cite{SayahianJahromi:2018irq}, the Barrow entropy \cite{Barrow:2020tzx}. 
the Kaniadakis entropy \cite{Kaniadakis:2005zk,Drepanou:2021jiv}, 
the Loop Quantum Gravity entropy \cite{Majhi:2017zao,Czinner:2015eyk,Mejrhit:2019oyi, Liu:2021dvj}, and etc. 
By applying the above entropies to the cosmology instead of the Bekenstein-Hawking entropy, the modifications of the FLRW equations can be obtained. 
{
Recently it has been shown that the entropic cosmology corresponding to various entropic functions is 
equivalent to holographic cosmology with suitable holographic cut-offs \cite{Nojiri:2021iko, Nojiri:2021jxf}. 
In this direction, the holographic energy density sourced from various entropies successfully describes the dark energy era 
of the universe, and generally known as the holographic dark energy (HDE) model 
\cite{Li:2004rb,Li:2011sd,Wang:2016och,Pavon:2005yx,Nojiri:2005pu,Gong:2004cb,Khurshudyan:2016gmb,Landim:2015hqa,
Gao:2007ep,Li:2008zq}. 
Besides the dark energy era, the holographic energy density is large during the early stage, which in turn triggers a viable inflationary scenario 
of the universe consistent with the Planck data 
\cite{Nojiri:2019kkp, Paul:2019hys, Bargach:2019pst, Elizalde:2019jmh, Oliveros:2019rnq}, 
and more interestingly, the holographic cosmology can unify early inflation to the late dark energy era in a covariant formalism \cite{Nojiri:2020wmh}.} 

In \cite{Howfundamental}, as generalizations, the authors have considered the three-parameter entropy-like quantity 
\begin{align}
\label{general6}
 S_\mathrm{G} \left( \alpha, \beta, \gamma \right)
= \frac{1}{\gamma} \left[ \left( 1 + \frac{\alpha}{\beta} S_\mathrm{BH} 
\right)^\beta - 1 \right] \,,
\end{align}
where $S_\mathrm{BH}$ represents the Bekenstein-Hawking entropy and is given by $S_\mathrm{BH} = \pi R_\mathrm{h}^2$, here $R_\mathrm{h}$ is the radius of the horizon under consideration. The entropic parameters $ \left( \alpha , \beta, \gamma \right)$ in the above expression are considered to be positive. Further, in \cite{Nojiri:2022dkr}, a four-parameters generalized entropy has been proposed, given by,
\begin{align}
\label{general1}
S_\mathrm{G} \left( \alpha_+ , \alpha_- , \beta , \gamma \right)
= \frac{1}{\gamma}\left[\left(1 + \frac{\alpha_+}{\beta}~S_\mathrm{BH}\right)^{\beta} 
 - \left(1 + \frac{\alpha_-}{\beta}~S_\mathrm{BH}\right)^{-\beta}\right] \,,
\end{align}
where all the parameters $\left( \alpha_+ , \alpha_- , \beta , \gamma \right) $ are positive.
In some suitable limits of the parameter values, the above entropies (\ref{general6}) and (\ref{general1}) reproduce the known entropies, 
that is, the Tsallis entropy, the R{\'e}nyi entropy, the Sharma-Mittal entropy, the Barrow entropy, the Kaniadakis entropy, and 
the Loop Quantum Gravity entropy. For example, the entopy in Eq.~(\ref{general1}) reduces to the known entropies for the following limits of the parameters \cite{Nojiri:2022dkr}: (a) to the Tsallis entropy for $\alpha_+ \rightarrow \infty$ and $\alpha_- = 0$, (b) to the R{\'e}nyi entropy for $\alpha_- = 0$, $\beta \rightarrow 0$, $\alpha_+ = \gamma$ and $\frac{\alpha_+}{\beta } \rightarrow \mathrm{finite}$, (c) to the Sharma-Mittal entropy for $\alpha _- = 0$ and $\alpha_+ = \gamma$ (d) to the Kaniadakis entropy for
$\beta \rightarrow \infty$ and $\alpha_+ = \alpha_- = \frac{\gamma}{2}$, (e) to the Loop Quantum Gravity entropy for $\alpha_- = 0$, $\beta \rightarrow \infty$ and
$\alpha_+ = \gamma$, respectively.\\

All of the above entropies have the following properties: 

\begin{enumerate}

\item {\it Generalized third law:} All these entropies vanish when the 
Bekenstein-Hawking entropy vanishes. In the third law of standard 
thermodynamics for closed systems in thermodynamic equilibrium, the 
quantity $\e^{S_\mathrm{BH}}$ expresses the number of 
states, or the volume of these states, and therefore the entropy 
$S_\mathrm{BH}$ vanishes when the temperature does because the ground (vacuum) state 
should be unique. By contrast, the Bekenstein-Hawking 
entropy $S_\mathrm{BH}$ diverges when the temperature $T$ vanishes and it goes 
to zero at infinite temperature. However, requiring any generalized 
entropy to vanish when the Bekenstein-Hawking entropy $S_\mathrm{BH}$ vanishes 
could be a natural requirement.

\item {\it Monotonically increasing functions:} All the above entropies 
are monotonically increasing functions of the Bekenstein-Hawking entropy 
$S_\mathrm{BH}$.

\item {\it Positivity:} All the above entropies are positive, as is the 
Bekenstein-Hawking entropy. This is natural because 
$\e^{S_\mathrm{BH}}$ corresponds to the number of states (or to the volume of 
these states), which is greater than unity.

\item {\it Bekenstein-Hawking limit:} All the above entropies reduce to 
the Bekenstein-Hawking entropy in an appropriate limit.

\end{enumerate}

{
In the context of entropic cosmology, it seems that for the Bekenstein-Hawking entropy, the relationship 
between the apparent horizon thermodynamics and the usual FLRW equations exists only for a special case, particularly when the matter content is considered 
to be a perfect fluid with EoS is given by $p = -\rho$ (where $p$ and $\rho$ represent the pressure and the energy density of the fluid, respectively) 
\cite{Cai:2005ra,Akbar:2006kj,Sanchez:2022xfh}. This raises the following important questions:
\begin{itemize}
\item Does there exist the relation between the thermodynamics of the apparent horizon and the usual FLRW equations for a general 
EoS of the matter fluid given by $p = w\rho$ where $w$ is constant EoS parameter of the fluid ? 
 
\item If yes, then what about this relationship 
when the EoS parameter is not a constant, but rather varies with the expansion of the universe?
 
\item How do the FLRW equations of the apparent horizon cosmology 
get modified for the known entropies proposed so far as well as for the generalized entropies mentioned above?
\end{itemize}

We will address these questions in this work.} 
In the next section, we try to clarify the relations between the thermodynamics of the apparent horizon and the FLRW equations. 
Then in section \ref{SecIII}, we consider several modifications of the entropy consistent with the FLRW equations. 
In section \ref{SecIV}, we start with the general definitions of entropy and we investigate how the FLRW equations 
are accordingly modified and in section \ref{SecV}, we discuss several interesting cosmological consequences of the modified FLRW equations. 
The last section is devoted to the summary and the conclusion.

\section{Thermodynamics of apparent horizon and FLRW equations}\label{SecII}
Here we briefly show, followed by \cite{Sanchez:2022xfh}, that for the Bekenstein-Hawking entropy, the apparent horizon thermodynamics leads to the usual FLRW equations only for the case when the normal matter content is described by the equation of state $p = -\rho$ where $p$ and $\rho$ represent the pressure and the energy density of the matter content respectively. By the ``usual FLRW equations'', we mean $H^2 + \frac{k}{a^2} = \frac{8\pi}{3}\rho$ and $\dot{H} - \frac{k}{a^2} = -4\pi\left(\rho + p\right)$ where $\rho$ and $p$ are contributed only from the normal matter content and the other quantities have standard meaning. Thus for a fluid $p = w\rho$ (with $w$ being the EoS parameter of the fluid), the apparent horizon thermodynamics based on the Bekenstein-Hawking entropy leads to the usual FLRW equations only for $w = -1$, which is the subject of the present section. This actually motivates us to construct an entropy function, based on which, the apparent horizon thermodynamics gives the usual FLRW equations for all possible values of $w$ including $w = -1$. We will consider this issue (with a general $w$) in Sec.~[\ref{SecIII}] where $w$ is considered to be a constant, and will also in Sec.~[\ref{SecIV}] where $w$ varies (i.e $w = w(t)$) with the universe's evolution.

We consider the FLRW universe, whose metric is given by,
\begin{align}
\label{dS7}
ds^2 = \sum_{\mu,\nu=0,1,2,3} g_{\mu\nu} dx^\mu dx^\nu = - dt ^2 + a( t )^2 \left( \frac{d r ^2}{1 - k r ^2} + r ^2 {d\Omega_2}^2 \right) \, ,
\end{align}
where $k$ can acquire $k=-1$, $0$, or $1$ describing the open (hyperboloid), flat (Eucledian) or closed (sphere) universe respectively. For $k = 0$, the effective energy density is equal to a critical value ($= \frac{3H_0^2}{8\pi G}$ where $H_0$ is the Hubble parameter at present era) and thus the universe is spatially flat; for $k = +1$, the effective energy density ($\rho_\mathrm{eff}$) is high enough compared to the critical density ($\rho_\mathrm{cr}$) and the universe is described as being a closed universe; while for $k = -1$, the $\rho_\mathrm{eff} < \rho_\mathrm{cr}$ holds so that the gravitational attraction is not sufficient to stop the expansion and thus the universe expands forever. We also define
\begin{align}
\label{dS7B}
d{s_\perp}^2 = \sum_{M,N=0,1} h_{\mu\nu} dx^M dx^N = - dt ^2 + \frac{a( t )^2 d r ^2}{1 - k r ^2} \, .
\end{align}
The radius of the apparent horizon $R_\mathrm{h}=R\equiv a(t)r$ for the FLRW universe is given by the solution of the equation 
$h^{MN} \partial_M R \partial_N R = 0$ (see \cite{Cai:2005ra,Akbar:2006kj,Sanchez:2022xfh}) which immediately leads to, 
\begin{align}
\label{dS14A}
R_\mathrm{h}=\frac{1}{\sqrt{ H^2 + \frac{k}{a^2}}}\, , 
\end{align}
with $H\equiv \frac{1}{a}\frac{da}{d t }$ represents the Hubble parameter of the universe. 
The surface gravity $\kappa$ on the apparent horizon is defined as \cite{Cai:2005ra}
\begin{align}
\label{SG3}
\kappa= \left. \frac{1}{2\sqrt{-h}} \partial_M \left( \sqrt{-h} h^{MN} \partial_N R \right) \right|_{R=R_\mathrm{h}}\, .
\end{align}
For the metric of Eq.~(\ref{dS7}), we have $R=a r $ and obtain 
\begin{align}
\label{SG2}
\kappa = - \frac{1}{R_\mathrm{h}} \left\{ 1 + \frac{{R_\mathrm{h}}^2}{2} \left( \dot H - \frac{k}{a^2} \right) \right\} \, ,
\end{align}
where the following expression is used, 
\begin{align}
\label{dS14AB}
\dot R_\mathrm{h} = - H {R_\mathrm{h}}^3 \left( \dot H - \frac{k}{a^2} \right) \, ,
\end{align}
The surface gravity of Eq.~(\ref{SG2}) is related with the temperature via $T_\mathrm{h} = \kappa/(2\pi)$, i.e 
\begin{align}
\label{AH2}
T_\mathrm{h} \equiv \frac{\left| \kappa \right|}{2\pi} 
= \frac{1}{2\pi R_\mathrm{h}} \left| 1 - \frac{\dot R_\mathrm{h}}{2 H R_\mathrm{h}} \right|
= \frac{1}{2\pi R_\mathrm{h}} \left| 1 + \frac{{R_\mathrm{h}}^2}{2} 
\left( \dot H - \frac{k}{a^2} \right) \right|\, .
\end{align}

The variation of the energy inside the apparent horizon is given by 
\begin{align}
\label{AH3}
dE= \frac{d}{dt} \left( \frac{4\pi}{3} {R_\mathrm{h}}^3 \rho \right) = \left( \frac{4\pi}{3} {R_\mathrm{h}}^3 \dot \rho + 4\pi \rho {R_\mathrm{h}}^2 \dot R_\mathrm{h} \right) dt\, .
\end{align}
Here $\rho$ is the energy density, and $p$ is the pressure. 
The thermodynamical relation, particularly the first law of thermodynamics gives 
\begin{align}
\label{AH3B}
dE = dQ - p dV = dQ- 4\pi p {R_\mathrm{h}}^2 \dot R_\mathrm{h} dt\, .
\end{align}
Let us assume that the entropy is given by the standard Bekenstein-Hawking relation, 
\begin{align}
\label{AH6}
S_\mathrm{BH} = \frac{A}{4}=\pi {R_\mathrm{h}}^2 \, .
\end{align}
In this note, we choose $G=\frac{\kappa^2}{8\pi}=1$. 
By using the standard thermodynamical relation, 
\begin{align}
\label{AH7}
TdS = dQ\, ,
\end{align}
and by identifying $T=T_\mathrm{h}$ and $S=S_\mathrm{BH}$, we find 
\begin{align}
\label{AH8}
\left| 1 + \frac{{R_\mathrm{h}}^2}{2} 
\left( \dot H - \frac{k}{a^2} \right) \right| 
H {R_\mathrm{h}}^3 \left( \dot H - \frac{k}{a^2} \right) 
= \frac{4\pi}{3} {R_\mathrm{h}}^3 \dot\rho 
+ 4\pi \left( p + \rho \right) H {R_\mathrm{h}}^5 \left( \dot H - \frac{k}{a^2} \right) \, .
\end{align}
By using the conservation law of the matter content, 
\begin{align}
\label{AH4}
0=\dot\rho + 3 H \left( \rho + p \right) \, ,
\end{align}
we rewrite Eq.~(\ref{AH8}) as follows, 
\begin{align}
\label{AH8BB}
\left| 1 + \frac{{R_\mathrm{h}}^2}{2} \left( \dot H - \frac{k}{a^2} \right) \right| H \left( \dot H - \frac{k}{a^2} \right) 
= \frac{4\pi}{3} \dot\rho \left( 1 - {R_\mathrm{h}}^2 \left( \dot H - \frac{k}{a^2} \right) \right) \, .
\end{align}
As shown in \cite{Sanchez:2022xfh}, Eq.~(\ref{AH8}) becomes consistent with the FLRW equations, 
\begin{align}
\label{AH9}
H^2 + \frac{k}{a^2} =&\, \frac{1}{{R_\mathrm{h}}^2} = \frac{8\pi}{3}\rho\, , \\
\label{AH10}
\left( \dot H - \frac{k}{a^2} \right)=&\, - 4\pi \left( \rho + p \right) = \frac{8\pi}{3}\dot \rho \, ,
\end{align}
{
only when $p = -\rho$, i.e the matter content is given by a perfect fluid having EoS parameter $= -1$. 

There seem to be some relations between the thermodynamics of the apparent horizon and the FLRW equations but 
we have found that the relationship exists only for a special case, $p=-\rho$. 
The reason could be because we naively considered the standard Bekenstein-Hawking entropy in Eq.~(\ref{AH6}). 
Motivated by this argument and to include the case $p \neq -\rho$, 
we consider the modification of the Bekenstein-Hawking entropy. In particular, we intend to construct an entropy that leads to 
the usual FLRW equations for a general EoS of the fluid given by $p = w\rho$ (with $w$ being the EoS parameter of the fluid). 
This is the subject of the next section.} 

\section{Entropy corresponding to general FLRW equations}\label{SecIII}

Here we intend to construct the modification of the Bekenstein-Hawking entropy to get the usual FLRW equations for a general EoS $p = w\rho$ where
$w$ is a constant (the case with varying $w$ is considered in the next section). If the modified entropy
is symbolized by $S$, then the thermodynamical relation Eq.~(\ref{AH7}) leads to, 
\begin{align}
\label{AH11}
 - \frac{1}{2\pi R_\mathrm{h}} \left| 1 + \frac{{R_\mathrm{h}}^2}{2} \left( \dot H - \frac{k}{a^2} \right) \right| 
\dot S = \frac{4\pi}{3} {R_\mathrm{h}}^3 \dot\rho \left( 1 - {R_\mathrm{h}}^2 \left( \dot H - \frac{k}{a^2} \right) \right) \, .
\end{align}
By using Eq.~(\ref{AH10}), we can rewrite Eq.~(\ref{AH11}) as follows, 
\begin{align}
\label{AH12}
 - \frac{1}{2\pi R_\mathrm{h}} \left( 1 + \frac{{R_\mathrm{h}}^2}{2} \left( \dot H - \frac{k}{a^2} \right) \right)
\dot S = H {R_\mathrm{h}}^3 \left( \dot H - \frac{k}{a^2} \right) \left( 1 - {R_\mathrm{h}}^2 \left( \dot H - \frac{k}{a^2} \right) \right) \, ,
\end{align}
where we assume $1 + \frac{{R_\mathrm{h}}^2}{2} \left( \dot H - \frac{k}{a^2} \right) > 0$. 
Further by using Eq.~(\ref{dS14AB}), we equivalently express the above equation as follows: 
\begin{align}
\label{AH13}
\left( 1 - \frac{\dot R_\mathrm{h}}{2H R_\mathrm{h}} \right) 
\dot S = 2\pi R_\mathrm{h} \dot R_\mathrm{h} \left( 1 + \frac{\dot R_\mathrm{h}}{H R_\mathrm{h}} \right) \, ,
\end{align}
on integrating which, we obtain
\begin{align}
\label{AH14}
S = \pi{R_\mathrm{h}}^2 + 6 \pi \int^t dt \frac{R_\mathrm{h} {\dot R_\mathrm{h}}^2}{2 H R_\mathrm{h} - \dot R_\mathrm{h}} \, .
\end{align}
The second term in Eq.~(\ref{AH14}) expresses the correction for the Bekenstein-Hawking entropy for the apparent horizon of Eq.~(\ref{AH6}). 
The second term is given by a function of the cosmological time $t$. 
By solving the expression of $R_\mathrm{h}=R_\mathrm{h}(t)$ with respect to $t$ and eliminating $t$ in the second term of Eq.~(\ref{AH14}), 
the second term can be expressed as a function of $R_\mathrm{h}$. 

It is often convenient to use the $e$-foldings $N$ defined by $a=\e^N$ instead of the cosmological time. 
Due to $\frac{d}{dt} = H \frac{d}{dN}$, Eq.~(\ref{AH14}) can be rewritten as 
\begin{align}
\label{AH14B}
S = \pi{R_\mathrm{h}}^2 + 6 \pi \int^N dN \frac{R_\mathrm{h} \left( \frac{d R_\mathrm{h}}{dN} \right)^2}{2 R_\mathrm{h} - \frac{d R_\mathrm{h}}{dN}} \, .
\end{align}
As an example, we consider the perfect fluid with a constant equation of state parameter $w$, $p=w\rho$, as the matter. 
Then the conservation law of Eq.~(\ref{AH4}) tells $\rho=\rho_0 \e^{-3\left(1+w\right)N}$ with a constant $\rho_0$. 
Then the first FLRW equation (\ref{AH9}) results to, 
\begin{eqnarray}
\label{w1}
R_\mathrm{h}= \sqrt{\frac{3}{8\pi \rho_0}} \e^{\frac{3}{2} \left(1+w\right) N}~~.
\end{eqnarray}
Clearly for $w = -1$, the radius of the apparent horizon becomes constant and thus the integral in Eq.~(\ref{AH14B}) vanishes. With the above expression of $R_\mathrm{h}$, Eq.~(\ref{AH14B}) leads to the desired entropy function as,
\begin{eqnarray}
\label{AH14BBN}
S = \pi{R_\mathrm{h}}^2 + \frac{27\left(1+w\right)^2}{\left( 1 - 3 w \right)}\left(\frac{3}{8\rho_0}\right) \int^N dN \e^{3\left(1+w\right)N}~~.
\end{eqnarray}
Performing the integration in Eq.~(\ref{AH14BBN}) and using $S_\mathrm{BH} = \pi R_\mathrm{h}^2$, we finally get
\begin{eqnarray}
S = \left[ 1 + \frac{9\left(1+w\right)}{\left( 1 - 3 w \right)} \right] S_\mathrm{BH}~~.
\label{AH14BB}
\end{eqnarray}
It is evident that for $w \neq -1$, the modified entropy ($S$: that produces the usual FLRW equations) gets a correction term over the Bekenstein-Hawking entropy; while for $w = -1$, the correction term vanishes and thus the entropy $S$ becomes similar to the Bekenstein-Hawking entropy. The latter statement is however expected from the demonstration of the previous section which clearly reveals that for the case $w = -1$, the
Bekenstein-Hawking entropy indeed leads to the usual FLRW equations. Furthermore it deserves mentioning that due to the presence of the factor $(1-3w)$ in the denominator of $S$, the modified entropy diverges when $w = \frac{1}{3}$ which corresponds to the radiation. This argues either of the following points: (1) there might not exist a suitable entropy which leads to the usual FLRW equations for the case $w = \frac{1}{3}$ i.e when the matter is only radiation, or, (2) there exists a more fundamental entropy which can produce the usual FLRW equations for all possible EoS of the matter content including $w = \frac{1}{3}$, while our constructed entropy in the present work is a sub-class (represented by $w \neq \frac{1}{3}$) of such fundamental entropy. However irrespective of these arguments, we hope that we make important progress to address the well defined question: ``Does there exist any entropy function that is able to give the usual FLRW equations for the whole spectrum of EoS of the matter content under consideration?''



Thus as a whole, we find the expression of the modified entropy in Eq.~(\ref{AH14BB}) that results in the usual FLRW equations for
general EoS given by $p = w\rho$ with $w \neq \frac{1}{3}$.
Especially, when the matter is given by a single perfect fluid with a constant equation of the state parameter, the entropy is different from
the Bekenstein-Hawking entropy by a factor as shown in Eq.~(\ref{AH14BB}).
In the next section, we will extend such development of the modified entropy to the case where the EoS parameter of the matter content
is not a constant, but rather varies with the cosmological evolution of the universe -- this makes the treatment more general.

\section{Modification of FLRW equations corresponding to general entropies}\label{SecIV}

We may consider the case that the entropy $S$ is a function of the Bekenstein-Hawking entropy for the apparent horizon in Eq.~(\ref{AH6}), 
$S=S\left( S_\mathrm{BH} = \pi {R_\mathrm{h}}^2 \right)$. 
Then Eq.~(\ref{AH11}) gives 
\begin{align}
\label{AH15}
\left( 1 + \frac{{R_\mathrm{h}}^2}{2} \left( \dot H - \frac{k}{a^2} \right) \right) H \left( \dot H - \frac{k}{a^2} \right) S'(S_\mathrm{BH}) 
= \frac{4\pi}{3} \dot\rho \left( 1 - {R_\mathrm{h}}^2 \left( \dot H - \frac{k}{a^2} \right) \right) \, ,
\end{align}
where the overprime symbolizes the derivative with respect to $S_\mathrm{BH}$ and we consider 
$1 + \frac{{R_\mathrm{h}}^2}{2} \left( \dot H - \frac{k}{a^2} \right) > 0$, again. 
By using the conservation law of the matter section, we can rewrite Eq.~(\ref{AH15}) as follows, 
\begin{align}
\label{AH16}
\left( 1 + \frac{\frac{{3R_\mathrm{h}}^2}{2} \left( \dot H - \frac{k}{a^2} \right)}{1 - {R_\mathrm{h}}^2 \left( \dot H - \frac{k}{a^2} \right)} 
\right) S' \left( \dot H - \frac{k}{a^2} \right) 
= -4\pi \left(\rho + p \right) \, ,
\end{align}
which can be regarded as the correction to the second FLRW equation. 
Correspondingly, the correction to the first FLRW equation can be obtained by the integration of the above equation, and is given by, 
\begin{align}
\label{AH17B}
\int dN \left( 1 + \frac{\frac{{3R_\mathrm{h}}^2}{2} \left( H \frac{dH}{dN} - \frac{k}{a^2} \right)}
{1 - {R_\mathrm{h}}^2 \left( H \frac{dH}{dN} - \frac{k}{a^2} \right)} 
\right) S' \frac{d}{dN}\left( H^2 + \frac{k}{a^2} \right) 
= \frac{8\pi}{3} \rho \, ,
\end{align}
where we use the e-folding number defined by $\frac{dN}{dt} = H$. The above equation 
can be further rewritten by using $S_\mathrm{BH} = \pi R_\mathrm{h}^2 = \pi/\left(H^2 + \frac{k}{a^2}\right)$, 
\begin{align}
\label{AH17E}
 - \int dN \left( 1 - \frac{\frac{3}{4S_\mathrm{BH}} \frac{dS_\mathrm{BH}}{dN}}
{1 + \frac{1}{2 S_\mathrm{BH}} \frac{dS_\mathrm{BH}}{dN}} 
\right) \left( \frac{1}{{S_\mathrm{BH}}^2} \right) \frac{dS_\mathrm{BH}}{dN} S'\left( S_\mathrm{BH} \right)
= \frac{8}{3} \rho \, .
\end{align}
If $\rho$ is given in terms of the $e$-folding $N$, $\rho=\rho(N)$, Eq.~(\ref{AH17E}) gives the differential equation
for $S_\mathrm{BH}$, 
\begin{align}
\label{AH17F}
 - \left( 1 - \frac{\frac{3}{4S_\mathrm{BH}} \frac{dS_\mathrm{BH}}{dN}}
{1 + \frac{1}{2 S_\mathrm{BH}} \frac{dS_\mathrm{BH}}{dN}} 
\right) \left( \frac{1}{{S_\mathrm{BH}}^2} \right) \frac{dS_\mathrm{BH}}{dN} S'\left( S_\mathrm{BH} \right)
= \frac{8}{3} \frac{d\rho}{dN} \, ,
\end{align}
which we will try to solve later. 

Eq,~(\ref{AH16}) suggests that in order to reproduce the standard FLRW equations, i.e $H^2 + \frac{k}{a^2} = \frac{8\pi}{3}\rho$ and $\dot{H} - \frac{k}{a^2} = -4\pi\left(\rho + p\right)$, we need to require
\begin{eqnarray}
 \left( \frac{1 + \frac{{R_\mathrm{h}}^2}{2} \left( \dot H - \frac{k}{a^2} \right)}{1 - {R_\mathrm{h}}^2 \left( \dot H - \frac{k}{a^2} \right)} 
\right) S'(S_\mathrm{BH}) = 1~,
\label{new1}
\end{eqnarray}
where recall that the overprime denotes the derivative with respect to $S_\mathrm{BH}$. On integrating the above equation,
\begin{eqnarray}
 S\left(S_\mathrm{BH}\right) = \int dS_\mathrm{BH} \left[ \frac{1 - {R_\mathrm{h}}^2 \left( \dot H - \frac{k}{a^2} \right)}{1 + \frac{{R_\mathrm{h}}^2}{2} \left( \dot H - \frac{k}{a^2} \right)}\right]~~.
 \label{New2}
\end{eqnarray}
Since we are dealing to reproduce the FLRW equations, we can use $H^2 + \frac{k}{a^2} = \frac{8\pi}{3}\rho = \frac{1}{R_\mathrm{h}^2}$ and $\dot{H} - \frac{k}{a^2} = -4\pi\left(\rho + p\right)$ in the above expression to get,
\begin{eqnarray}
 S = \int dS_\mathrm{BH} \left[1 + \frac{9\left(1+w(t)\right)}{\left(1-3w(t)\right)}\right]~~,
 \label{AH21}
\end{eqnarray}
where $w(t)$ represents the EoS of the matter content, i.e $p = w(t)\rho$. We now individually discuss two cases, particularly when $w$ is a constant or when
$w(t)$ varies with the cosmological expansion of the universe. First case: For constant $w$, Eq.~(\ref{AH21}) immediately leads to the entropy: $S = \left[1 + \frac{9\left(1+w\right)}{\left(1-3w\right)}\right]S_\mathrm{BH}$, which is similar to Eq.~(\ref{AH14BB}), as expected. Second case: For a time varying $w(t)$,
Eq.(\ref{AH21}) cannot be realized without specifying the scale factor $a=a(t)$ as a function of the cosmological time
$t$ because we assume $S$ is a function of 
the Bekenstein-Hawking entropy $S_\mathrm{BH}$ and therefore $S$ only depends on $R_\mathrm{h}$ but not on $\dot H$. Thus if
we consider a specific scale factor $a(t)$ given by a function $f(t)$ of the cosmological time $t$, $a(t)=f(t)$,
$R_\mathrm{h}$ is also given 
by a function of $t$ via Eq.~(\ref{dS14A}), $R_\mathrm{h}=R_\mathrm{h}(t)=\frac{a(t)}{\sqrt{{\dot a(t)}^2 + k}}$, which can be solved with respect 
to $t$ as a function of $R_\mathrm{h}$, $t=t\left( R_\mathrm{h} \right)$. 
In this case, $\dot H$ is also given by a function of $R_\mathrm{h}$, i.e $\dot{H} = \dot{H}(R_\mathrm{h}) = 
\dot{H}\left(\sqrt{S_\mathrm{BH}/\pi}\right)$. Owing to which, Eq.(\ref{AH21}) can be integrated to give $S$,
\begin{align}
\label{AH21-N}
S \left( S_\mathrm{BH} \right) = \int dS_\mathrm{BH}
\left[\frac{1 - \frac{S_\mathrm{BH}}{\pi} \left( \dot H - \frac{k}{a^2} \right)}
{1 + \frac{S_\mathrm{BH}}{2\pi} \left( \dot H - \frac{k}{a^2} \right)}\right] \, ,
\end{align}
with $\dot{H} = \dot{H}\left(\sqrt{S_\mathrm{BH}/\pi}\right)$ and $a = a\left(\sqrt{S_\mathrm{BH}/\pi}\right)$. 
Because Eq.~(\ref{AH16}) reduces to the standard FLRW equation provided Eq.~(\ref{AH21}) is satisfied,
$\rho$ might have an exotic behavior given by $\rho=\rho(t) \equiv \frac{3}{8\pi} \left( H(t)^2 + \frac{k}{a(t)^2} \right)$ in general, 
which can be realized by an exotic equation of state.\\

Having obtained the modified FLRW equations as in Eq.~(\ref{AH16}) and Eq.~(\ref{AH17E}) 
by starting with a general entropy, we now find the explicit form of the modification for several kinds 
of definitions of entropy. As examples, we may consider the Tsallis entropy \cite{tsallis} (see also \cite{Ren:2020djc,Nojiri:2019skr}), 
the R{\'e}nyi entropy (\cite{renyi}, see also \cite{Nishioka:2014mwa, Czinner:2015eyk,Tannukij:2020njz,Promsiri:2020jga,Samart:2020klx}) and the 
generalized entropies \cite{Howfundamental,Nojiri:2022dkr}. 

The Tsallis entropy appears in the non-extensive statistics as appears in the systems with long-range interactions, 
\begin{align}
\label{TS1}
S_\mathrm{T} = S_0 \left( \frac{S_\mathrm{BH}}{S_0} 
\right)^\delta \,,
\end{align}
where $S_0$ and $\delta$ are constants. 
The parameter $\delta$ quantifies the non-extensivity. 
The standard Bekenstein-Hawking entropy $S_\mathrm{BH}$ in Eq.~(\ref{AH6}) is recovered 
in the limit of $\delta \to 1$. 

The R{\'e}nyi entropy is defined as 
\begin{align}
\label{RS1}
S_\mathrm{R}=\frac{1}{\alpha} \ln \left( 1 + \alpha S_\mathrm{BH} \right) \, ,
\end{align}
which contains a parameter $\alpha$. 
Originally the R\'enyi entropy was proposed to be an index specifying the amount of information. 

In \cite{Howfundamental}, we have considered the three-parameter entropy-like quantity given in Eq.~(\ref{general6}). 
By writing $\gamma = \left( \alpha/\beta \right)^{\beta} $, the limit $\alpha\to \infty$ yields 
\begin{align}
\label{general7}
\lim_{\alpha\to \infty} S_\mathrm{G} \left( \alpha, \beta, 
\gamma=\left( \frac{\alpha}{\beta} \right)^\beta \right)
= {S_\mathrm{BH}}^\beta \, .
\end{align}
The choice $\beta=\delta$ gives the Tsallis entropy shown in Eq.~(\ref{TS1}). 
On the other hand, if we consider the limit in which $\alpha\rightarrow 0 $ and 
$\beta \rightarrow 0$ simultaneously while keeping $ \alpha/\beta $ 
finite, we obtain the R{\'e}nyi entropy in Eq.~(\ref{RS1}) by replacing $\alpha/\beta$ 
with $\alpha$ and choosing $\gamma=\alpha$, 
\begin{align}
\label{general8}
 S_\mathrm{G} \left( \alpha\to 0, \beta\to 0, \gamma; 
\frac{\alpha}{\beta} = \mbox{finite} \right)
\to \frac{1}{\gamma} \ln \left( 1 + \frac{\alpha}{\beta} \, S_\mathrm{BH} 
\right) = \frac{1}{\alpha}\ln \left( 1 + \alpha S_\mathrm{BH} \right) \,.
\end{align}
By considering other limits, the three-parameter entropy-like quantity of Eq.~(\ref{general6}) also reproduces 
the Sharma-Mittal entropy \cite{SayahianJahromi:2018irq} and the Barrow entropy \cite{Barrow:2020tzx}. 

In \cite{Nojiri:2022dkr} we introduced four-parameter dependent generalized entropy construct as shown in Eq.~(\ref{general1}), 
which reduces to the known entropies proposed so far including the Tsallis and the R\'{e}nyi ones in suitable limits of the parameters.

Having described various entropy functions including the generalized ones, we now determine the modified FLRW Eq.(\ref{AH16}) for some explicit 
forms of entropy. For the Tsallis entropy, Eq.~(\ref{AH16}) turns out to be, 
\begin{align}
\label{AH18}
\delta \left( 1 + \frac{\frac{{3R_\mathrm{h}}^2}{2} \left( \dot H - \frac{k}{a^2} \right)}{1 - {R_\mathrm{h}}^2 \left( \dot H - \frac{k}{a^2} \right)} 
\right) \left( \dot H - \frac{k}{a^2} \right) \left( \frac{\pi {R_\mathrm{h}}^2}{S_0} \right)^{\delta-1} 
= -4\pi \left(\rho + p \right) \, .
\end{align}
Eq.~(\ref{AH18}) indicates that the Tsallis entropy generates an effective energy density ($\rho_\mathrm{T}$) and an effective pressure ($p_\mathrm{T}$) given by:
\begin{eqnarray}
 \rho_\mathrm{T} + p_\mathrm{T} = \frac{1}{4\pi}\left(\dot{H} - \frac{k}{a^2}\right)
 \left\{\delta \left( 1 + \frac{\frac{{3R_\mathrm{h}}^2}{2} \left( \dot H - \frac{k}{a^2} \right)}{1 - {R_\mathrm{h}}^2 \left( \dot H - \frac{k}{a^2} \right)}
\right) \left( \frac{\pi {R_\mathrm{h}}^2}{S_0} \right)^{\delta-1}  - 1\right\}~~,
 \label{N1}
\end{eqnarray}
where recall that $R_\mathrm{h}^2 = \left(H^2 + \frac{k}{a^2}\right)^{-1}$. In the case of the R{\'e}nyi entropy, Eq.(\ref{AH16}) takes the following form,
\begin{eqnarray}
\label{AH19}
\left( 1 + \frac{\frac{{3R_\mathrm{h}}^2}{2} \left( \dot H - \frac{k}{a^2} \right)}{1 - {R_\mathrm{h}}^2 \left( \dot H - \frac{k}{a^2} \right)} 
\right) \left( \dot H - \frac{k}{a^2} \right) \frac{1}{1 + \alpha \pi {R_\mathrm{h}}^2}
= -4\pi \left(\rho + p \right)~~,
\end{eqnarray}
which leads to the energy density ($\rho_\mathrm{R}$) and pressure ($p_\mathrm{R}$) sourced by the R{\'e}nyi entropy as,
\begin{eqnarray}
 \rho_\mathrm{R} + p_\mathrm{R} = \frac{1}{4\pi}\left(\dot{H} - \frac{k}{a^2}\right)
 \left\{\frac{1}{1 + \alpha \pi {R_\mathrm{h}}^2}\left( 1 + \frac{\frac{{3R_\mathrm{h}}^2}{2} \left( \dot H - \frac{k}{a^2} \right)}{1 - {R_\mathrm{h}}^2 \left( \dot H - \frac{k}{a^2} \right)}
\right) - 1\right\}~~.
 \label{N2}
\end{eqnarray}
Moreover, in the case of the 4-parameter generalized entropy in Eq.~(\ref{general1}), the apparent horizon cosmology is governed by,
\begin{eqnarray}
\label{AH20}
\frac{1}{\gamma}
\left( 1 + \frac{\frac{{3R_\mathrm{h}}^2}{2} \left( \dot H - \frac{k}{a^2} \right)}{1 - {R_\mathrm{h}}^2 \left( \dot H - \frac{k}{a^2} \right)} 
\right)\left( \dot H - \frac{k}{a^2} \right) \left[\alpha_+\left( 1 + \frac{\alpha_+}{\beta} \pi {R_\mathrm{h}}^2 \right)^{\beta-1} +
\alpha_-\left( 1 + \frac{\alpha_-}{\beta} \pi {R_\mathrm{h}}^2 \right)^{-\beta-1}\right]= -4\pi \left(\rho + p \right)~.
\end{eqnarray}
This results to,
\begin{eqnarray}
 \rho_\mathrm{G} + p_\mathrm{G} = \frac{1}{4\pi}\left(\dot{H} - \frac{k}{a^2}\right)
 &\Bigg\{&\frac{1}{\gamma}
\left( 1 + \frac{\frac{{3R_\mathrm{h}}^2}{2} \left( \dot H - \frac{k}{a^2} \right)}{1 - {R_\mathrm{h}}^2 \left( \dot H - \frac{k}{a^2} \right)}
\right)\times\nonumber\\
&\Big[&\alpha_+\left( 1 + \frac{\alpha_+}{\beta} \pi {R_\mathrm{h}}^2 \right)^{\beta-1} +
\alpha_-\left( 1 + \frac{\alpha_-}{\beta} \pi {R_\mathrm{h}}^2 \right)^{-\beta-1}\Big] - 1\Bigg\}
 \label{N3}
\end{eqnarray}
generated from the 4-parameter generalized entropy. The presence of such energy densities sourced from the respective entropies indeed modify the field equations and lead to some interesting cosmological phenomena in the context of entropic cosmology, which we will discuss in the next section.

In summary of this section, we have explicitly found how the FLRW equations are modified according to the definitions in entropies 
by choosing the Tsallis entropy (see Eq.~(\ref{TS1})), the R{\'e}nyi entropy (see Eq.~(\ref{RS1})) and the three-parameter entropy-like quantity 
(see Eq.~(\ref{general6})), as examples.

\section{Cosmological consequences of the modified FLRW equations}\label{SecV}

Recall that the modified FLRW equation corresponding to a general form of entropy of the apparent horizon cosmology is determined in Eq.~(\ref{AH17F}). 
We intend to solve this equation and examine its various consequences in this section. First, we consider the case without matter $\rho=0$, which gives 
\begin{align}
\label{AH22}
0=\frac{dS_\mathrm{BH}}{dN} \quad \mbox{or} \quad 
0= 4S_\mathrm{BH} - \frac{dS_\mathrm{BH}}{dN} \, .
\end{align}
The situation does not depend on the choice of $S=S\left( S_\mathrm{BH} \right)$. 
The solution of the first equation is $S_\mathrm{BH}=\mathrm{constant}$ and therefore $R_\mathrm{h}=\mathrm{constant}$, which give the 
effective cosmological constant, 
\begin{align}
\label{AH23}
\frac{1}{{R_\mathrm{h}}^2} = H^2 + \frac{k}{a^2} = \Lambda\, .
\end{align}
The second equation in Eq.~(\ref{AH22}) gives $S_\mathrm{BH} \propto \e^{4N}$, which gives 
\begin{align}
\label{AH23B}
\frac{1}{{R_\mathrm{h}}^2} = H^2 + \frac{k}{a^2} = C \e^{-4N}\, , 
\end{align}
with a constant $C$, which corresponds to an exotic matter with the equation of state parameter $w=3$. 
Therefore even in the case that there is no cosmological constant or real matter, Eq.~(\ref{AH17F}) generates a universe with cosmological 
constant or an exotic matter. 

Because it is difficult to solve the non-linear differential equation (\ref{AH17F}) in general, we assume
$\left| \frac{1}{S_\mathrm{BH}} \frac{dS_\mathrm{BH}}{dN} \right| < 1$ and expand Eq.~(\ref{AH17F}) with
respect to $\frac{1}{S_\mathrm{BH}} \frac{dS_\mathrm{BH}}{dN}$. The condition $\frac{1}{S_\mathrm{BH}} \frac{dS_\mathrm{BH}}{dN} < 1$ refers that the change of the Hubble parameter is rather slow, in particular $\frac{1}{H}\frac{dH}{dN} < 1$ which is well valid during the early stage as well as during the late stage of the universe. In this section, since we are mainly interested in explaining the inflation and the dark energy era by entropic cosmology, the aforementioned condition is suitable for our purpose. The expansion of Eq.~(\ref{AH17F}) with respect to $\frac{1}{S_\mathrm{BH}} \frac{dS_\mathrm{BH}}{dN}$ leads to,
\begin{align}
\label{AH24}
 - \left( \frac{S' \left( S_\mathrm{BH} \right)}{{S_\mathrm{BH}}^2} \right) \frac{dS_\mathrm{BH}}{dN} 
\sim \frac{8}{3} \frac{d\rho}{dN} 
 - \frac{3}{4S_\mathrm{BH}} \frac{dS_\mathrm{BH}}{dN} \left( \frac{S' \left( S_\mathrm{BH} \right)}{{S_\mathrm{BH}}^2} \right) 
\frac{dS_\mathrm{BH}}{dN} \left( S_\mathrm{BH} \right) 
\sim \frac{8}{3} \frac{d\rho}{dN} \left\{ 1 + \frac{3}{4S_\mathrm{BH}} \frac{dS_\mathrm{BH}}{dN} \right\} \, ,
\end{align}
First, we consider the case that the entropy is the Bekenstein-Hawking entropy $S=S_\mathrm{BH}$, 
\begin{align}
\label{AH25}
\frac{d}{dN}\left( \frac{1}{S_\mathrm{BH}} \right) 
\sim \frac{8}{3} \frac{d\rho}{dN} \left\{ 1 + \frac{3}{4S_\mathrm{BH}} \frac{dS_\mathrm{BH}}{dN} \right\} 
\sim \frac{8}{3} \frac{d\rho}{dN} \left\{ 1 - \frac{3}{4} \frac{d}{dN} \ln \left( \frac{8}{3} \left( \rho + C_1 \right) \right) \right\} \, .
\end{align}
As a matter, we consider the perfect fluid with a constant equation of state parameter $w$, $\rho=\rho_0 \e^{-3\left(1+w\right)N}$ 
and we obtain 
\begin{align}
\label{AH26}
\frac{d}{dN}\left( H^2 + \frac{k}{a^2} \right) 
\sim \frac{8}{3}\pi \left\{ \frac{d\rho}{dN} + \frac{27 (1+w)^2}{4} \frac{{\rho_0}^2\e^{-6\left(1+w\right)N}}{\rho_0 \e^{-3\left(1+w\right)N} + C_1} \right\} \, .
\end{align}
that is, 
\begin{align}
\label{AH27}
H^2 + \frac{k}{a^2} \sim&\, \frac{8}{3}\pi \left\{ \rho_0 \e^{-3\left(1+w\right)N} + C_1 
+ \frac{27 (1+w)^2}{4} \left( - \frac{\rho_0 \e^{-3\left(1+w\right)N}}{3(1+w)} + \frac{C_1}{3(1+w)} \ln \left( \rho_0 \e^{-3\left(1+w\right)N} + C_1 \right) \right)
\right\} \nonumber \\
=&\, \frac{8}{3}\pi \left\{ \left( 1 - \frac{9(1+w)}{4} \right) \rho_0 \e^{-3\left(1+w\right)N} + C_1 
+ \frac{9 (1+w) C_1}{4} \ln \left( \rho_0 \e^{-3\left(1+w\right)N} + C_1 \right) \right\} \, .
\end{align}
Therefore the contribution from the matter decreases by a factor, $1 - \frac{9(1+w)}{4}$ and there effectively appears a contribution 
from an exotic matter expressed in the last term of Eq.~(\ref{AH27}).

In the cases of the Tsallis entropy, the R{\'e}nyi entropy, the generalized entropy having 3 or 4 parameters,
Eq.~(\ref{AH17F}) gives the modification of the first FLRW equation 
even in the zeroth order of $\frac{1}{S_\mathrm{BH}} \frac{dS_\mathrm{BH}}{dN}$. 
\begin{itemize}
 \item \underline{For Tsallis entropy}: In the case of the Tsallis entropy, the zeroth order equation of Eq.~(\ref{AH17F}) has the following form,
\begin{align}
\label{AH28-0}
 - \delta {S_0}^{1-\delta} {S_\mathrm{BH}}^{\delta -3} \frac{dS_\mathrm{BH}}{dN} 
\sim \frac{8}{3} \frac{d\rho}{dN} \, ,
\end{align}
which can be integrated to give 
\begin{align}
\label{AH28}
 - \frac{\delta}{\delta - 2} {S_0}^{1-\delta} {S_\mathrm{BH}}^{\delta - 2} 
\sim \frac{8}{3} \left( \rho + \Lambda \right) \, .
\end{align}
Here $\Lambda$ appears as a constant of the integration. 
We rewrite Eq.~(\ref{AH28}) in the form similar to the first FLRW equation, 
\begin{align}
\label{AH29}
H^2 + \frac{k}{a^2} \sim \pi \left\{ \frac{8}{3}\frac{2 - \delta}{\delta} {S_0}^{\delta -1} \left( \rho + \Lambda \right) \right\}^{\frac{1}{2-\delta}} \, .
\end{align}
In this regard, we consider two cases, particularly $\rho = 0$ and $\rho \neq 0$ respectively.
\subsubsection*{Case-1: $\rho = 0$, i.e the normal matter is absent}
In this case, Eq.~(\ref{AH29}) takes the following form:
\begin{align}
\label{AH-new1}
H^2 + \frac{k}{a^2} \sim \pi \left\{ \frac{8}{3}\frac{2 - \delta}{\delta} {S_0}^{\delta -1} \Lambda \right\}^{\frac{1}{2-\delta}} \, .
\end{align}
Therefore when the normal matter content is absent, the entropic cosmology corresponds to the Tsallis entropy results to a constant energy density leading to an accelerating expansion of the universe. For suitable $\delta$ and $\Lambda$, the early or the late era of the universe can be described by such an accelerating expansion.
\subsubsection*{ Case-2: $\rho \neq 0$, i.e in presence of normal matter}
In the case with a perfect fluid having constant EoS parameter $w$, the energy density of the fluid goes as $\rho=\rho_0 \e^{-3\left(1+w\right)N}$, and thus Eq.~(\ref{AH29}) turns out to be,
\begin{eqnarray}
\label{AH-new2}
H^2 + \frac{k}{a^2} \sim \pi \left\{ \frac{8}{3}\frac{2 - \delta}{\delta} {S_0}^{\delta -1} \left( \rho_0 \e^{-3\left(1+w\right)N} + \Lambda \right) \right\}^{\frac{1}{2-\delta}}~~,
\end{eqnarray}
Here it deserves mentioning that for $\delta = 1$, the Tsallis entropy gets similar to the Bekenstein-Hawking entropy, in which case, Eq.~(\ref{AH-new2}) resembles with the usual FLRW equation. This is expected because for $S = S_\mathrm{BH}$, the condition in Eq.~(\ref{new1}) is automatically satisfied in the leading order of
$\left|\frac{1}{S_\mathrm{BH}}\frac{dS_\mathrm{BH}}{dN}\right|$. However for $\delta \neq 1$, the scenario gets modified compared to the usual FLRW equation, and the modification is given by the above equation which may lead to some interesting cosmology. For instance if $\Lambda = 0$ (i.e the bare cosmological constant coming from the integration constant in the governing equation), then the Hubble parameter of Eq.~(\ref{AH-new2}) is controlled by an effective EoS parameter given by,
\begin{eqnarray}
\label{general6D-TS}
w_\mathrm{eff} = - 1 + \frac{1+w}{2 - \delta}~~.
\end{eqnarray}
Clearly $w_\mathrm{eff} \neq -1$ and $w_\mathrm{eff} \neq w$ due to the presence of the entropic parameter $\delta$. We now show that the above $w_\mathrm{eff}$ may be consistent with the dark energy EoS parameter of the present universe. Considering $\delta = 0.1$ and $w = -0.2$, one gets $w_\mathrm{eff} = -0.957$ which is well consistent with the Planck data regarding the dark energy EoS parameter \cite{Planck:2018vyg}. However with $\delta = 1$, for which Eq.~(\ref{AH-new2}) produces the usual FLRW equation, the effective EoS parameter becomes $w_\mathrm{eff} = w$ from Eq.~(\ref{general6D-TS}) -- this indicates that for $w = -0.957 = w_\mathrm{eff}$, the model becomes compatible with the present dark energy era. Thereby it is clear that in order to have a consistent dark energy EoS parameter with the Planck data,
the matter content needs to be closer to the phantom structure for $\delta = 1$ compared to the case of $\delta \neq 1$. This reveals an advantage for considering the modified FLRW equation, sourced by the entropic cosmology corresponds to the Tsallis entropy, over the usual FLRW scenario.

\item \underline{For R{\'e}nyi entropy}: in the case of the R{\'e}nyi entropy, Eq.~(\ref{AH17F}) gives,
\begin{align}
\label{AH30}
 - \frac{1}{{S_\mathrm{BH}}^2} \frac{1}{1 + \alpha S_\mathrm{BH}} 
\frac{dS_\mathrm{BH}}{dN}
\sim \frac{8}{3} \frac{d\rho}{dN} \, ,
\end{align}
where we retain up-to the zeroth order of $\left|\frac{1}{S_\mathrm{BH}}\frac{ds_\mathrm{BH}}{dN}\right|$. Integrating the above equation, one obtains
\begin{align}
\label{AH31}
\frac{1}{S_\mathrm{BH}} + \alpha \ln \left| \frac{S_\mathrm{BH}}{1 + \alpha S_\mathrm{BH}} \right|
\sim \frac{8}{3} \left( \rho + \Lambda \right) \, .
\end{align}
There appears a constant of the integration $\Lambda$, again. During the universe's evolution, the condition $GH^2 \ll 1$ (or equivalently $H^2 \ll 1$ in the unit of $G = 1$ with $G$ being the Newton's gravitational constant) safely holds, thus $S_\mathrm{BH}$ is large. Owing to which, Eq.~(\ref{AH31}) can be solved, and is given by,
\begin{align}
\label{AH33}
H^2 + \frac{k}{a^2} \sim \frac{8}{3}\pi \left( \rho + \Lambda + \frac{3}{8}\alpha \ln \left| \alpha \right| \right)\, .
\end{align}
Therefore the entropic cosmology corresponds to the R{\'e}nyi entropy sources an effective energy density given by $\rho_\mathrm{R} = \frac{3}{8}\alpha \ln \left| \alpha \right|$.
\subsubsection*{Case-1: $\rho = 0$}
In this case, Eq.~(\ref{AH33}) becomes,
\begin{align}
\label{AH-new3}
H^2 + \frac{k}{a^2} \sim \frac{8}{3}\pi \left( \Lambda + \frac{3}{8}\alpha \ln \left| \alpha \right| \right) = \frac{8}{3}\pi\left(\Lambda + \rho_\mathrm{R}\right)\, .
\end{align}
Therefore even if the cosmological constant $\Lambda$ vanishes, $\Lambda=0$, the
$\rho_\mathrm{R} = \alpha \ln \left| \alpha \right|$ plays the role of the effective cosmological constant.
This may tell that the R{\'e}nyi entropic energy density may generate the late-time accelerating expansion of the universe provided
$\alpha \ln \left| \alpha \right|>0$.
\subsubsection*{Case-2: $\rho \neq 0$}
In presence of a normal matter content having EoS: $p = w\rho$ with $w$ being a constant, we have $\rho = \rho_0 \e^{-3\left(1+w\right)N}$ and consequently
Eq.~(\ref{AH33}) comes as,
\begin{eqnarray}
\label{AH-new4}
H^2 + \frac{k}{a^2} \sim \frac{8}{3}\pi \left( \rho_0 \e^{-3\left(1+w\right)N} + \Lambda + \frac{3}{8}\alpha \ln \left| \alpha \right| \right)~~,
\end{eqnarray}
where the effective energy density is contributed from the normal matter, the R{\'e}nyi entropic energy density and the cosmological constant, i.e
$\rho_\mathrm{eff} = \rho + \rho_\mathrm{R} + \Lambda$. This results to some non-trivial evolution of $H(N)$ for different values of $w$.

\item \underline{For the 3-parameter generalized entropy shown in Eq.~(\ref{general6})}: In this scenario, Eq.~(\ref{AH17F}) up-to the zeroth order has the following form,
\begin{align}
\label{AH34}
 - \frac{\alpha}{{\gamma S_\mathrm{BH}}^2} \left( 1 + \frac{\alpha}{\beta} S_\mathrm{BH} 
\right)^{\beta-1}\frac{dS_\mathrm{BH}}{dN}
\sim \frac{8}{3} \frac{d\rho}{dN} \, .
\end{align}
Due to the reason that $S_\mathrm{BH}$ is large, Eq.~(\ref{AH34}) is safely approximated as
\begin{align}
\label{AH37}
 - \frac{\alpha^\beta}{\gamma \beta^{\beta-1} } {S_\mathrm{BH}}^{\beta - 3} \frac{dS_\mathrm{BH}}{dN}
\sim \frac{8}{3} \frac{d\rho}{dN} \, ,
\end{align}
which in turn gives
\begin{align}
\label{AH38}
H^2 + \frac{k}{a^2} \sim \pi \left\{ \frac{8\gamma \beta^{\beta -1} \left(2 - \beta \right)}{3\alpha^\beta} 
\left( \rho + \Lambda \right) \right\}^{\frac{1}{2-\beta}} \, .
\end{align}
\subsubsection*{Case-1: In absence of normal matter}
In the case of $\rho = 0$, Eq.~(\ref{AH38}) becomes,
\begin{align}
\label{AH-new5}
H^2 + \frac{k}{a^2} \sim \pi \left\{ \frac{8\gamma \beta^{\beta -1} \left(2 - \beta \right)}{3\alpha^\beta} \Lambda \right\}^{\frac{1}{2-\beta}} \, .
\end{align}
This leads to a constant Hubble parameter and may describe the late phase of the universe. In particular,
by using Eq.~(\ref{AH-new5}), we find $H\sim \sqrt{G^{-1} \left( G^2 \Lambda \right)^\frac{1}{2 - \beta}}$ (here to clarify the dimension, we write $G$ explicitly).
We may choose, $\Lambda\sim \left( 10^{15}\,\mathrm{GeV} \right)^4$.
Then because $G\sim \left( 10^{19}\,\mathrm{GeV} \right)^{-2}$, the Hubble parameter is given by $H\sim 10^{19 - \frac{8}{2 - \beta}}\, \mathrm{GeV}= 10^{28 - \frac{8}{2 - \beta}}\, \mathrm{eV}$. Considering $\beta\sim 1.87$, we obtain $H\sim 10^{-33}\, \mathrm{eV}$, which corresponds to the value of the Hubble rate $H$ in the present universe. Therefore the entropic cosmology corresponds to the 3-parameter generalized entropy may describe the accelerating expansion of the present late universe.
\subsubsection*{Case-2: In presence of normal matter}
In this case, the energy density of the matter content behaves as $\rho = \rho_0 \e^{-3\left(1+w\right)N}$ where $w$ is the constant EoS of the matter. As a result, the evoluton of $H(N)$ is controlled by,
\begin{align}
\label{AH-new6}
H^2 + \frac{k}{a^2} \sim \pi \left\{ \frac{8\gamma \beta^{\beta -1} \left(2 - \beta \right)}{3\alpha^\beta}
\left( \rho_0 \e^{-3\left(1+w\right)N} + \Lambda \right) \right\}^{\frac{1}{2-\beta}} \, .
\end{align}
For instance, if the energy density coming from the normal matter is larger than that of $\Lambda$, then the Hubble parameter is dominated by an effective EoS parameter given by,
\begin{eqnarray}
\label{AH39}
w_\mathrm{eff} = - 1 + \frac{1+w}{2 - \beta}~~,
\end{eqnarray}
which is clearly different from the cosmological constant. Moreover Eq.~(\ref{AH39}) depicts that
$w_\mathrm{eff} \neq w$ due to the presence of the entropic parameter $\beta$. Thus the 3-parameter generalized entropy under consideration modifies the Hubble parameter, and hence the universe's evolution, compared to the usual scenario in presence of the normal matter.


\item \underline{For the 4-parameter generalized entropy shown in Eq.~(\ref{general1})}: Due to the reason that $S_\mathrm{BH} \sim \frac{\pi}{GH^2} \gg 1$, the
4-parameter generalized entropy of Eq.~(\ref{general1}) may be approximated as,
\begin{align}
\label{general6B}
S_\mathrm{G} \left( \alpha_{\pm}, \beta, \gamma \right) 
\sim \frac{1}{\gamma}\left(\frac{\alpha_+}{\beta}\right)^{\beta} {S_\mathrm{BH}}^{\beta} \,,
\end{align}
The above behavior of $S_\mathrm{G}$ is similar to that of
the Tsallis entropy by identifying ${S_0}^{1-\delta}\equiv \frac{1}{\gamma}\left(\frac{\alpha_+}{\beta}\right)^{\beta}$ and $\delta \equiv \beta$.
Therefore, similarly to (\ref{AH29}), the governing equation for the 4-parameter generalized entropy can be expressed as,
\begin{eqnarray}
\label{AH-new7}
H^2 + \frac{k}{a^2} \sim \pi \left\{ \frac{8}{3}\frac{\gamma\left(2 - \beta\right)}{\beta} \left(\frac{\beta}{\alpha_+}\right)^{\beta} \left( \rho + \Lambda \right) \right\}^{\frac{1}{2-\beta}}~~,
\end{eqnarray}
which, due to $\rho_0 \e^{-3\left(1+w\right)N}$, is equivalently expressed by,
\begin{eqnarray}
 \label{AH-new8}
H^2 + \frac{k}{a^2} \sim \pi \left\{ \frac{8}{3}\frac{\gamma\left(2 - \beta\right)}{\beta} \left(\frac{\beta}{\alpha_+}\right)^{\beta}
\left( \rho_0 \e^{-3\left(1+w\right)N} + \Lambda \right) \right\}^{\frac{1}{2-\beta}}~~.
\end{eqnarray}
Clearly for $\beta = 1$ and $\gamma = \alpha_+$, the above equation resembles with the usual FLRW equation. This is expected because for $\beta = 1$ and $\gamma= \alpha_+$, the geeralized entropy in Eq.~(\ref{general6B}) gets similar to the Bekenstein-Hawking entropy and thus the condition in Eq.~(\ref{new1}) is automatically satisfied in the leading order of $\left|\frac{1}{S_\mathrm{BH}}\frac{dS_\mathrm{BH}}{dN}\right|$. However for $\beta \neq 1$, Eq.~(\ref{AH-new8}) provides a modified cosmological scenario which results to some interesting phenomena. For instance -- (1) if $\Lambda \gg \rho$, then Eq.~(\ref{AH-new8}) results to a de-Sitter expansion of the universe which can describe the early phase of the universe, or (2) if $\rho \gg \Lambda$, then the Hubble parameter of Eq.~(\ref{AH-new8}) is controlled by an effective EoS parameter given by,
\begin{eqnarray}
\label{general6D}
w_\mathrm{eff} = -1 + \frac{1+w}{2 - \beta}~~,
\end{eqnarray}
which is equivalent to that of in the case of Tsallis entropy with the replacement $\delta \equiv \beta$ (see Eq.~(\ref{general6D-TS})). Therefore the advantage for using the Tsallis entropy demonstrated after Eq.~(\ref{general6D-TS}) is similarly shared by the entropic cosmology corresponds to the 4-parameter generalized entropy.
\end{itemize}

Before concluding, we may note that there could be other kinds of thermal effects as discussed in \cite{Capozziello:2021bpk}.
Eq.~(\ref{AH2}) tells that the temperature $T_\mathrm{h}$ becomes large when $R_\mathrm{h}$ is small. 
Because Eq.~(\ref{dS14A}) tells that $R_\mathrm{h}$ is small when $H$ is large and $a$ is small, 
in the early universe, we have the high temperature $T_\mathrm{h}$. 
We may expect the generation of thermal radiation as in the case of the Hawking radiation. 
Because the thermal radiation is proportional to the fourth power of the temperature, 
the energy density $\rho$ could be modified as 
\begin{align}
\label{TH1}
\rho \to \rho + \alpha_0 {T_\mathrm{h}}^4 
=&\, \rho + \frac{\alpha_0}{\left(2\pi R_\mathrm{h}\right)^4} \left\{ 1 + \frac{{R_\mathrm{h}}^2}{2} 
\left( \dot H - \frac{k}{a^2} \right) \right\}^4 \nonumber \\
=&\, \rho + \frac{\alpha_0}{\left(2\pi \right)^4} \left( H^2 + \frac{k}{a^2} \right)^2
\left\{ 1 + \frac{1}{2\left( H^2 + \frac{k}{a^2} \right)} 
\left( \dot H - \frac{k}{a^2} \right) \right\}^4 \nonumber \\
\sim &\, \rho + \frac{\alpha_0}{\left(2\pi \right)^4} \left( H^2 + \frac{k}{a^2} \right)^2\, ,
\end{align}
where $\alpha_0$ is a constant and in the last line, we have assumed $H$ is large and $a$ is small, again.
Clearly Eq.~(\ref{TH1}) gives a certain modification of the FLRW equation.
For example, Eq.~(\ref{AH-new7}) regarding the 4-parameter entropy can be expressed as,
\begin{eqnarray}
\label{TH2}
H^2 + \frac{k}{a^2} \sim \pi \left\{ \frac{8}{3(2\pi)^4}\frac{\gamma\alpha_0\left(2 - \beta\right)}{\beta} \left(\frac{\beta}{\alpha_+}\right)^{\beta} \left(H^2 + \frac{k}{a^2}\right)^2\right\}^{\frac{1}{2-\beta}}~~,
\end{eqnarray}
which gives 
\begin{align}
\label{TH3}
H^2 + \frac{k}{a^2} \sim 0\, , \quad \mbox{or} \quad H^2 + \frac{k}{a^2} \sim \left(\frac{1}{\pi}\right)^{\frac{2-\beta}{\beta}}\left\{\frac{3\left(2\pi \right)^4\beta}{8\alpha_0 \gamma(2-\beta)}\left(\frac{\alpha_+}{\beta}\right)^{\beta}\right\}^{\frac{1}{\beta}} \, .
\end{align}
The first equation in Eq.~(\ref{TH3}) may give a rather trivial solution but in the second equation, 
the effective cosmological constant $\Lambda_\mathrm{eff} = \left(\frac{1}{\pi}\right)^{\frac{2-\beta}{\beta}}\left\{\frac{3\left(2\pi \right)^4\beta}{8\alpha_0 \gamma(2-\beta)}\left(\frac{\alpha_+}{\beta}\right)^{\beta}\right\}^{\frac{1}{\beta}}$ appears,
which may generate inflation. 
These results do not change qualitatively even if we consider the Bekenstein-Hawking entropy instead of 
the three-parameter entropy-like quantity because the three-parameter entropy-like quantity is proportional to the 
Bekenstein-Hawking entropy when the Bekenstein-Hawking entropy is small, which is always true in the early universe. 
Many other definitions of the entropies also have the property that the entropies reduce to the 
Bekenstein-Hawking entropy when the Bekenstein-Hawking entropy is small and therefore we obtain similar results. 
If there is still a non-linearity when the Bekenstein-Hawking entropy is small, we may obtain some specific results.

\section{Summary}\label{SecVII}
In the thermodynamic context of the apparent horizon, the Bekenstein-Hawking entropy is related to the usual FLRW equations
only for a special case, particularly when the normal matter content is considered to be a perfect fluid with the EoS given by $p = -\rho$ (with $p$ and
$\rho$ are the pressure and the energy density of the fluid under consideration, respectively). 
By the ``usual FLRW equations'', we mean $H^2 + \frac{k}{a^2} = \frac{8\pi}{3}\rho$ and $\dot{H} - \frac{k}{a^2} = -4\pi\left(\rho + p\right)$ where $\rho$ and $p$ are contributed only from the normal matter content and the other quantities have standard meaning. Thus for a fluid $p = w\rho$ (with $w$ being the EoS parameter of the fluid), the apparent horizon thermodynamics based on the Bekenstein-Hawking entropy leads to the usual FLRW equations only for $w = -1$.
Motivated by this argument and to include the case $p \neq -\rho$, we search for a suitable modification of the Bekenstein-Hawking entropy.
Accordingly, we develop a modified entropy, based on which, the apparent horizon thermodynamics gives the usual FLRW equations for all possible values of $w$ including $w = -1$. Such development of the modified entropy is also extended to the case when the EoS parameter
is not a constant but rather varies with the cosmological evolution of the universe. 
The newly developed entropy acquires a correction over the Bekenstein-Hawking entropy and shares the following properties -- 
(1) monotonically increases with the Bekenstein-Hawking entropy ($S_\mathrm{BH}$), 
(2) vanishes in the limit $S_\mathrm{BH} \rightarrow 0$ and
(3) converges to $S_\mathrm{BH}$ for $w = -1$ that recovers the aforementioned case of the Bekenstein-Hawking entropy. Here it deserves mentioning that
the modified entropy that we have constructed diverges for $w = \frac{1}{3}$ which corresponds to the radiation. This argues either of the following points: (1) there might not exist a suitable entropy which leads to the usual FLRW equations for the case $w = \frac{1}{3}$ i.e when the matter is only radiation, or, (2) there exists a more fundamental entropy which can produce the usual FLRW equations for all possible EoS of the matter content including $w = \frac{1}{3}$, while our constructed entropy in the present work is a sub-class (represented by $w \neq \frac{1}{3}$) of such fundamental entropy. However irrespective of these arguments, we hope that we make important progress to address the well defined question: ``Does there exist any entropy function that is able to give the usual FLRW equations for the whole spectrum of EoS of the matter content under consideration?'' One more point we would like to mention is that the modified entropy consistent with the usual FLRW equations differs from all the known entropies like the Tsallis, R\'{e}nyi, Barrow, Sharma-Mittal, Kaniadakis, and Loop Quantum Gravity entropies proposed so far.
Owing to this argument, we investigate how the FLRW equations of the apparent horizon cosmology get corrected for a general form of entropy 
and discuss its possible consequences. 
It turns out that the entropy sources some effective energy density that leads to some interesting cosmology during the early and late stages of 
the universe. 
For example -- (1) the entropic cosmology corresponds to the R{\'e}nyi entropy sources an effective cosmological constant which may generate the late time acceleration of the universe even in the absence of bare cosmological constant coming from the integration constant in the governing equation,
(2) in a more general set-up, if one considers the four-parameter generalized entropy proposed in \cite{Nojiri:2022dkr}, the model can successfully describe the dark energy EoS in consistent with the Planck data. In particular, the effective EoS parameter corresponds to the 4-parameter generalized entropy comes as
$w_\mathrm{eff} = - 1 + \frac{1+w}{2 - \beta}$ where $w$ is the constant EoS of the normal matter and $\beta$ symbolizes the entopic parameter. Clearly $w_\mathrm{eff} \neq -1$ and $w_\mathrm{eff} \neq w$ due to the presence of the entropic parameter $\beta$. Therefore with $\beta = 0.1$ and $w = -0.2$, one gets an observationally viable dark energy EoS parameter in respect to the Planck data. However with $\beta = 1$, for which the generalized entropy reduces to the Bekenstein-Hawking entropy and produces usual FLRW equation, the effective EoS parameter becomes $w_\mathrm{eff} = w$ -- this indicates that for $w = -0.957 = w_\mathrm{eff}$, the model becomes compatible with the present dark energy era. Thereby it is clear that in order to have a consistent dark energy EoS parameter with the Planck data,
the matter content needs to be closer to the phantom structure for $\beta = 1$ compared to the case of $\beta \neq 1$. This reveals an advantage for considering the modified FLRW equation, sourced by the entropic cosmology corresponds to the 4-parameter generalized entropy ($S_\mathrm{G}$), over the usual FLRW scenario.
the acceleration during the early stage as well as during the late stage of the universe may be explained in a unified manner. Such advantage or success for using the 4-parameter generalized entropy is also shared by the Tsallis entropy or even by the 3-parameter generalized entropy, as demonstrated in Sec~[\ref{SecV}]. Furthermore, generalized entropy construct applied to the apparent horizon may lead to an even more deep and more interesting
relationship between the fundamental entropy and gravity.

\section*{Acknowledgments}
This work was supported by MINECO (Spain), project PID2019-104397GB-I00 (SDO). This work was partially supported by the program Unidad de Excelencia
MarÃ­a de Maeztu CEX2020-001058-M (SDO). This research was also supported in part by the
International Centre for Theoretical Sciences (ICTS) for the online program - Physics of the Early Universe (code: ICTS/peu2022/1) (TP).


\begin{thebibliography}{99}

\bibitem{Bekenstein:1973ur}
J.~D.~Bekenstein,
Phys. Rev. D \textbf{7} (1973), 2333-2346
doi:10.1103/PhysRevD.7.2333

\bibitem{Hawking:1975vcx}
S.~W.~Hawking,
Commun. Math. Phys. \textbf{43} (1975), 199-220
[erratum: Commun. Math. Phys. \textbf{46} (1976), 206]
doi:10.1007/BF02345020

\bibitem{Jacobson:1995ab} 
T.~Jacobson,
Phys.\ Rev.\ Lett.\ {\bf 75}, 1260 (1995)
[gr-qc/9504004].

\bibitem{Padmanabhan:2003gd} 
T.~Padmanabhan,
Phys.\ Rept.\ {\bf 406}, 49 (2005)
[gr-qc/0311036].

\bibitem{Padmanabhan:2009vy} 
T.~Padmanabhan,
Rept.\ Prog.\ Phys.\ {\bf 73}, 046901 (2010)
[arXiv:0911.5004 [gr-qc]].


\bibitem{Hayward:1997jp}
S.~A.~Hayward,
Class. Quant. Grav. \textbf{15} (1998), 3147-3162
doi:10.1088/0264-9381/15/10/017
[arXiv:gr-qc/9710089 [gr-qc]].

\bibitem{Cai:2005ra}
R.~G.~Cai and S.~P.~Kim,
JHEP {\bf 0502}, 050 (2005)
[arXiv:hep-th/0501055].


\bibitem{Akbar:2006kj} 
M.~Akbar and R.~G.~Cai,
Phys.\ Rev.\ D {\bf 75}, 084003 (2007)
[hep-th/0609128].

\bibitem{Cai:2006rs} 
R.~G.~Cai and L.~M.~Cao,
Phys.\ Rev.\ D {\bf 75}, 064008 (2007)
[gr-qc/0611071].





\bibitem{Akbar:2006er} 
M.~Akbar and R.~G.~Cai,
Phys.\ Lett.\ B {\bf 635}, 7 (2006)
[hep-th/0602156].









\bibitem{Paranjape:2006ca} 
A.~Paranjape, S.~Sarkar and T.~Padmanabhan,
Phys.\ Rev.\ D {\bf 74}, 104015 (2006)
[hep-th/0607240].

\bibitem{Sheykhi:2007zp}
A.~Sheykhi, B.~Wang and R.~G.~Cai,
Nucl.\ Phys.\ B {\bf 779}, 1 (2007)
[arXiv:hep-th/0701198].



\bibitem{Jamil:2009eb} 
M.~Jamil, E.~N.~Saridakis and M.~R.~Setare,
Phys.\ Rev.\ D {\bf 81}, 023007 (2010)
[arXiv:0910.0822 [hep-th]].

\bibitem{Cai:2009ph} 
R.~G.~Cai and N.~Ohta,
Phys.\ Rev.\ D {\bf 81}, 084061 (2010)
[arXiv:0910.2307 [hep-th]].

\bibitem{Wang:2009zv} 
M.~Wang, J.~Jing, C.~Ding and S.~Chen,
Phys.\ Rev.\ D {\bf 81}, 083006 (2010)
[arXiv:0912.4832 [gr-qc]].

\bibitem{Jamil:2010di} 
M.~Jamil, E.~N.~Saridakis and M.~R.~Setare,
JCAP {\bf 1011}, 032 (2010)
[arXiv:1003.0876 [hep-th]].

\bibitem{Gim:2014nba} 
Y.~Gim, W.~Kim and S.~H.~Yi,
JHEP {\bf 1407}, 002 (2014)
[arXiv:1403.4704 [hep-th]].

\bibitem{Fan:2014ala} 
Z.~Y.~Fan and H.~Lu,
Phys.\ Rev.\ D {\bf 91}, no. 6, 064009 (2015)
[arXiv:1501.00006 [hep-th]].



\bibitem{DAgostino:2019wko}
R.~D'Agostino,
Phys.\ Rev.\ D {\bf 99} (2019) no.10, 103524
[arXiv:1903.03836 [gr-qc]].





\bibitem{Sanchez:2022xfh}
L.~M.~Sanchez and H.~Quevedo,
[arXiv:2208.05729 [gr-qc]].




\bibitem{tsallis}
C.~Tsallis, C, 
Journal of Statistical Physics. 52 (1-2) (1988), 479-487 
doi:10.1007/BF01016429

\bibitem{Ren:2020djc}
J.~Ren,
JHEP \textbf{05} (2021), 080
doi:10.1007/JHEP05(2021)080
[arXiv:2012.12892 [hep-th]].

\bibitem{Nojiri:2019skr}
S.~Nojiri, S.~D.~Odintsov and E.~N.~Saridakis,
Eur. Phys. J. C \textbf{79} (2019) no.3, 242
doi:10.1140/epjc/s10052-019-6740-5
[arXiv:1903.03098 [gr-qc]].

\bibitem{renyi}
A.~R{\'{e}}nyi, 
Proceedings of the Fourth Berkeley Symposium on Mathematics, Statistics 
and Probability, University of California Press (1960), 547-56

\bibitem{Nishioka:2014mwa}
T.~Nishioka,
JHEP \textbf{07} (2014), 061
doi:10.1007/JHEP07(2014)061
[arXiv:1401.6764 [hep-th]].

\bibitem{Czinner:2015eyk}
V.~G.~Czinner and H.~Iguchi,
Phys. Lett. B \textbf{752} (2016), 306-310
doi:10.1016/j.physletb.2015.11.061
[arXiv:1511.06963 [gr-qc]].

\bibitem{Tannukij:2020njz}
L.~Tannukij, P.~Wongjun, E.~Hirunsirisawat, T.~Deesuwan and C.~Promsiri,
Eur. Phys. J. Plus \textbf{135} (2020) no.6, 500
doi:10.1140/epjp/s13360-020-00517-2
[arXiv:2002.00377 [gr-qc]].

\bibitem{Promsiri:2020jga}
C.~Promsiri, E.~Hirunsirisawat and W.~Liewrian,
Phys. Rev. D \textbf{102} (2020) no.6, 064014
doi:10.1103/PhysRevD.102.064014
[arXiv:2003.12986 [hep-th]].

\bibitem{Samart:2020klx}
D.~Samart and P.~Channuie,
[arXiv:2012.14828 [hep-th]].




\bibitem{SayahianJahromi:2018irq}
A.~Sayahian Jahromi, S.~A.~Moosavi, H.~Moradpour, J.~P.~Morais Gra\c{c}a, I.~P.~Lobo, I.~G.~Salako and A.~Jawad,
Phys. Lett. B \textbf{780} (2018), 21-24
doi:10.1016/j.physletb.2018.02.052
[arXiv:1802.07722 [gr-qc]].

\bibitem{Barrow:2020tzx}
J.~D.~Barrow,
Phys. Lett. B \textbf{808} (2020), 135643
doi:10.1016/j.physletb.2020.135643
[arXiv:2004.09444 [gr-qc]].


\bibitem{Kaniadakis:2005zk}
G.~Kaniadakis,
Phys. Rev. E \textbf{72} (2005), 036108
doi:10.1103/PhysRevE.72.036108
[arXiv:cond-mat/0507311 [cond-mat]].

\bibitem{Drepanou:2021jiv}
N.~Drepanou, A.~Lymperis, E.~N.~Saridakis and K.~Yesmakhanova,
[arXiv:2109.09181 [gr-qc]].


\bibitem{Majhi:2017zao}
A.~Majhi,
Phys. Lett. B \textbf{775}, 32-36 (2017)
doi:10.1016/j.physletb.2017.10.043
[arXiv:1703.09355 [gr-qc]].

\bibitem{Mejrhit:2019oyi}
K.~Mejrhit and S.~E.~Ennadifi,
Phys. Lett. B \textbf{794}, 45-49 (2019)
doi:10.1016/j.physletb.2019.03.055

\bibitem{Liu:2021dvj}
Y.~Liu,
doi:10.1209/0295-5075/ac3f52
[arXiv:2112.15077 [gr-qc]].



\bibitem{Nojiri:2021iko}
S.~Nojiri, S.~D.~Odintsov and T.~Paul,
Symmetry \textbf{13} (2021) no.6, 928
doi:10.3390/sym13060928
[arXiv:2105.08438 [gr-qc]].

\bibitem{Nojiri:2021jxf}
S.~Nojiri, S.~D.~Odintsov and T.~Paul,
Phys. Lett. B \textbf{825} (2022), 136844
doi:10.1016/j.physletb.2021.136844
[arXiv:2112.10159 [gr-qc]].

\bibitem{Li:2004rb}
M.~Li,
Phys.\ Lett.\ B {\bf 603} (2004) 1
doi:10.1016/j.physletb.2004.10.014
[hep-th/0403127].

\bibitem{Li:2011sd}
M.~Li, X.~D.~Li, S.~Wang and Y.~Wang,
Commun. Theor. Phys. \textbf{56} (2011), 525-604
doi:10.1088/0253-6102/56/3/24
[arXiv:1103.5870 [astro-ph.CO]].

\bibitem{Wang:2016och}
S.~Wang, Y.~Wang and M.~Li,
Phys.\ Rept.\ {\bf 696} (2017) 1
doi:10.1016/j.physrep.2017.06.003
[arXiv:1612.00345 [astro-ph.CO]].

\bibitem{Pavon:2005yx}
D.~Pavon and W.~Zimdahl,
Phys.\ Lett.\ B {\bf 628} (2005) 206
doi:10.1016/j.physletb.2005.08.134
[gr-qc/0505020].

\bibitem{Nojiri:2005pu}
S.~Nojiri and S.~D.~Odintsov,
Gen.\ Rel.\ Grav.\ {\bf 38} (2006) 1285
doi:10.1007/s10714-006-0301-6
[hep-th/0506212].



\bibitem{Gong:2004cb}
Y.~g.~Gong, B.~Wang and Y.~Z.~Zhang,
Phys.\ Rev.\ D {\bf 72} (2005) 043510
doi:10.1103/PhysRevD.72.043510
[hep-th/0412218].


\bibitem{Khurshudyan:2016gmb}
M.~Khurshudyan,
Astrophys. Space Sci. \textbf{361} (2016) no.12, 392
doi:10.1007/s10509-016-2981-z

\bibitem{Landim:2015hqa}
R.~C.~G.~Landim,
Int.\ J.\ Mod.\ Phys.\ D {\bf 25} (2016) no.04, 1650050
doi:10.1142/S0218271816500504
[arXiv:1508.07248 [hep-th]].

\bibitem{Gao:2007ep}
C.~Gao, F.~Wu, X.~Chen and Y.~G.~Shen,
Phys.\ Rev.\ D {\bf 79} (2009) 043511
doi:10.1103/PhysRevD.79.043511
[arXiv:0712.1394 [astro-ph]].

\bibitem{Li:2008zq}
M.~Li, C.~Lin and Y.~Wang,
JCAP {\bf 0805} (2008) 023
doi:10.1088/1475-7516/2008/05/023
[arXiv:0801.1407 [astro-ph]].


\bibitem{Nojiri:2019kkp}
S.~Nojiri, S.~D.~Odintsov and E.~N.~Saridakis,
Phys. Lett. B \textbf{797} (2019), 134829
doi:10.1016/j.physletb.2019.134829
[arXiv:1904.01345 [gr-qc]].

\bibitem{Paul:2019hys}
T.~Paul,
EPL \textbf{127} (2019) no.2, 20004
doi:10.1209/0295-5075/127/20004
[arXiv:1905.13033 [gr-qc]].

\bibitem{Bargach:2019pst}
A.~Bargach, F.~Bargach, A.~Errahmani and T.~Ouali,
Int. J. Mod. Phys. D \textbf{29} (2020) no.02, 2050010
doi:10.1142/S0218271820500108
[arXiv:1904.06282 [hep-th]].

\bibitem{Elizalde:2019jmh}
E.~Elizalde and A.~Timoshkin,
Eur. Phys. J. C \textbf{79} (2019) no.9, 732
doi:10.1140/epjc/s10052-019-7244-z
[arXiv:1908.08712 [gr-qc]].

\bibitem{Oliveros:2019rnq}
A.~Oliveros and M.~A.~Acero,
EPL \textbf{128} (2019) no.5, 59001
doi:10.1209/0295-5075/128/59001
[arXiv:1911.04482 [gr-qc]].

\bibitem{Nojiri:2020wmh}
S.~Nojiri, S.~D.~Odintsov, V.~K.~Oikonomou and T.~Paul,
Phys. Rev. D \textbf{102} (2020) no.2, 023540
doi:10.1103/PhysRevD.102.023540
[arXiv:2007.06829 [gr-qc]].

\bibitem{Howfundamental}
S.~Nojiri, S.~D.~Odintsov and V.~Faraoni,
Phys. Rev. D \textbf{105} (2022) no.4, 044042
doi:10.1103/PhysRevD.105.044042
[arXiv:2201.02424 [gr-qc]].



\bibitem{Nojiri:2022dkr}
S.~Nojiri, S.~D.~Odintsov and T.~Paul,
Phys. Lett. B \textbf{831} (2022), 137189
doi:10.1016/j.physletb.2022.137189
[arXiv:2205.08876 [gr-qc]].

\bibitem{Capozziello:2021bpk}
S.~Capozziello, S.~Nojiri and S.~D.~Odintsov,
Phys. Dark Univ. \textbf{33} (2021), 100867
doi:10.1016/j.dark.2021.100867
[arXiv:2104.10936 [gr-qc]].


\bibitem{Planck:2018vyg}
N.~Aghanim \textit{et al.} [Planck],
Astron. Astrophys. \textbf{641} (2020), A6
[erratum: Astron. Astrophys. \textbf{652} (2021), C4]
doi:10.1051/0004-6361/201833910
[arXiv:1807.06209 [astro-ph.CO]].


\end{thebibliography}
\end{document}